# A potential biomarker of androgen exposure in European bullhead (*Cottus sp.*) kidney

Mélanie Villeret[1,2], Sabrina Jolly[1,3], Laure Wiest[4], Emmanuelle Vulliet[4], Anne Bado-Nilles[1,3], Jean-Marc Porcher[1], Stéphane Betoulle[3], Christophe Minier[2], Wilfried Sanchez[1*]

1. Institut National de l'Environnement Industriel et des Risques (INERIS), Unité d'écotoxicologie in vitro et in vivo, BP2, 60550, Verneuil-en-Halatte, France

2. Université du Havre, Laboratoire écotoxicologie et milieux aquatiques, BP540, 76058 Le Havre, France

3. Université de Reims Champagne-Ardenne, Unité de Recherche Vignes et Vins de Champagne - Stress et Environnement EA 2069, Laboratoire Ecologie-Ecotoxicologie, BP 1039, 51687 Reims Cedex 2, France

4. Service Central d'Analyse du CNRS - USR59, Echangeur de Solaize, Chemin du Canal, 69360 Solaize, France

* Corresponding author. Tel : +33 (0)3 44 61 81 21 ; Fax: +33 (0)3 44 55 67 67
E-mail address: Wilfried.Sanchez@ineris.fr (W. Sanchez)

**Abstract**

The aim of this study was to identify a signal that could be used as an androgen exposure indicator in the European bullhead (*Cottus sp.*). For this purpose, the ultra-structure of the kidney was characterized to identify normal structure of this organ, and histological changes previously described in the kidney of breeding male bullheads were quantified using the Kidney Epithelium Height (KEH) assay previously developed and validated for the stickleback. In the next step, the effect of trenbolone acetate (TbA), a model androgen, was assessed to identify potential androgenic regulation of bullhead kidney hypertrophy. Measurement of KEH performed on adult non-breeding male and female bullheads exposed for 14 and 21 days to 0, 1.26 and 6.50 µg/L showed that kidney hypertrophy is induced in a dose-dependent manner, confirming the hypothesis that the European bullhead possesses a potential biomarker of androgen exposure. Combined with the wide distribution of the European bullhead in European countries and the potential of this fish species for environmental toxicology studies in field and laboratory conditions, the hypothesis of a potential biomarker of androgen exposure offers interesting perspectives for the use of the bullhead as a relevant sentinel fish species in monitoring studies. Inducibility was observed with high exposure concentrations of TbA. Further studies are needed to identify molecular signals that could be more sensitive than KEH.

# 1. Introduction

The kidney of fish is a composite organ involved in three physiological functions: excretory, haematopoietic and endocrine. Classically, the kidney of teleost fish is divided into two parts. The head kidney, located in the anterior part of the organ, is integrated in the endocrine system and is involved in the stress response mediated by the hypothalamus-pituitary-interrenal cell axis and by the hypothalamus-sympathetic nervous system-chromaffin tissue axis (Donaldson 1981; Wendelar Bonga 1997). The trunk kidney located in the posterior area of the organ derives from mesonephros and it is heavily involved in excretory and osmo-regulatory functions (Hinton et al. 1992).

Due to the major role of the kidney in fish physiology, histopathological analysis of this organ is widely used to assess chronic adverse effects of environmental stressors including chemical exposure in wild fish. Around the world, several authors report histological alterations, for example cellular hypertrophy, vacuolation or degeneration, hemorrhage, enlargement or dilation of the glomerulus, in the kidneys of fish from contaminated sites (Silva and Martinez 2007; Ayas et al. 2007; Pacheco and Santos 2002). Hence, histological analysis of fish kidney is considered as a good means of evaluating the level of chronic stress to which a fish population is exposed, that may be used as an effect-based monitoring tool (Hinton et al. 1992). In the three-spined stickleback (*Gasterosteus aculeatus*), histological analysis of the kidney is a means of assessing androgen exposure. Indeed, the kidney of the stickleback hypertrophies after exposure to androgenic chemicals, when it starts to produce spiggin, a glue protein involved in the building of the nest (Katsiadaki et al. 2002). This hypertrophy can be quantified using the Kidney Epithelium Height (KEH) measurement previously developed by (Borg 1993) or the quantification of spiggin production using an ELISA (Katsiadaki et al. 2002; Sanchez et al. 2008a). Due to this physiological property, the stickleback is considered as a relevant model fish species to assess the (anti)androgenic effects of endocrine-disrupting chemicals (EDCs) in laboratory and field studies (Sanchez et al. 2008b; Katsiadaki et al. 2006; Pettersson et al. 2007; Wartman et al. 2009). Chemical analysis has led to the identification of chemicals involved in this biological activity and showed that natural androgenic compounds such as androstenedione, equol, dehydrotestosterone or androsterone represent a large part of measured activity (Hill et al. 2010; Thomas et al. 2002; Kinani et al. 2009). Field studies have described physiological changes associated with androgen exposure such as spiggin induction in stickleback from contaminated sites (Sanchez et al. 2008b; Katsiadaki et al. 2006; Pettersson et al. 2007; Wartman et al. 2009) or elongated anal fin rays, resembling the male gonopodium, in female mosquitofish (*Gambusia sp.*) collected downstream of a pulp mill effluent discharge (Bortone et al. 1989; Howell and Denton 1989). In this context, it is relevant to identify and develop novel indicators complementarily to stickleback spiggin and mosquitofish gonopodium, to address the contamination of European water bodies by androgens and associated effects.

The present work was designed to identify a signal that could be used as an androgen exposure indicator in the European bullhead (*Cottus sp.*). Previous work has revealed gender-dependent hypertrophy due to an increase of nephritic channel size in breeding male bullheads. This phenomenon is comparable to the natural change of stickleback kidney ultra-structure associated with spiggin synthesis during the breeding season (Bucher and Hofer 1993; Hentschel 1982). For this purpose, the ultra-structure of the kidney was characterized and sexual dimorphism was quantified using the Kidney Epithelium Height (KEH) assay previously developed and validated for the stickleback (Borg 1993). In the next step, the effect of trenbolone acetate (TbA), a model androgen, was assessed to identify potential androgenic regulation of bullhead kidney hypertrophy.

## 2. Material and methods

### 2.1. Chemicals

Trenbolone acetate (17β-Acetoxyestra-4,9,11-trien-3-one) and Dimethylsulfoxyde (DMSO) were purchased from Sigma-Aldrich (Saint Quentin, France) and were of analytical grade (≥98 % purity).

## 2.2. Fish origin and maintenance

To characterize sexual dimorphism and seasonal variability of KEH, male and female adult bullheads (mean weight = 6.22 ± 3.46 g; mean length = 75.94 ± 11.89 mm) were electrofished during March (n=15) and July 2011 (n=20), at an uncontaminated site located at the Arré river (Avrechy, Oise, France). These fish were immediately sacrificed.

Bullheads (mean weight = 2.4 ± 1.1 g; mean length = 59.1 ± 7.7 mm) were sampled from an uncontaminated site located at the Avre river (Davenescourt, Somme, France) during March 2010 and were used to characterize TbA exposure. These fish were transported to the laboratory (INERIS, Verneuil en Halatte, France) and maintained for two months in a flow-through tank (600 L, 8.8 ± 0.2 mg/L $O_2$, 350 µS/cm). During this period, the light/dark cycle was natural, the water temperature was 15 ± 2°C and the fish were fed daily with frozen red chironomid larvae.

## 2.3. Chemical exposure

Prior to exposures, the bullheads were randomly distributed in 5 litre tanks. All the fish were acclimated to experimental semi-static conditions for 1 week. During this acclimatization phase, the water temperature was 17.5 °C ± 1°C and the photoperiod was natural. The water was totally renewed every day and the fish were fed every other day with frozen red chironomid larvae.

The bullheads were exposed to two concentrations of TbA (approximately 1 and 5 µg/L) for 14 or 21 days in similar conditions to the acclimatization week. For each sampling time and concentration, fish were randomly collected for histological analysis.

TbA was dissolved in dimethylsulfoxyde (DMSO) and added to the experimental water with a maximum final concentration below 0.01%. For each exposure time, a solvent control group containing only DMSO was performed.

## 2.4. Tissue sampling and histological analysis

Fish were sacrificed weighed and measured. The gonads and kidneys were dissected, weighed and fixed in Bouin's solution for a minimum of 24h. The tissues were then dehydrated through a graded series of ethanol, cleared with toluene, and embedded in paraffin. Sections (5 µm) were mounted on glass slides and stained with hematoxylin and eosin. The gonad slide examination allowed gender confirmation. The examination of kidney histological sections allowed the calculation of kidney epithelium cell height (KEH) according to the method described by Borg (Borg 1993) and validated for the bullhead. Briefly, the height of 10 secondary proximal epithelium cells was measured using a Leica DMIL (Leica, Melville, NY, USA) phase contrast microscope equipped with a Qicam Fast 1394 camera (Qimaging, Canada) and analysis of pictures was performed with Qcapture Software (Qimaging, Canada). A KEH mean was calculated for each examined fish.

## 2.5. Water sampling and chemical analysis

The TbA concentration in experimental water was determined by chemical analysis. For each concentration tested, each week, 10 mL was taken from each condition just after the renewal of water over a period of 3 weeks, pooled (3 x 10 mL) in a glass bottle and stored at -20°C prior to chemical analysis. Analyses of chemicals were performed by liquid chromatography coupled with tandem mass spectrometry (LC-MS/MS). Liquid chromatography was performed on a HP1290 HPLC system (Agilent Technologies). The chromatographic separations were performed on an Eclipse Plus C18 RRHD (50 x 2.1 mm, 1.8 µm) column from Agilent Technologies. For the analysis, the column oven temperature was set to 60°C; the injection volume was 10 µL. The mobile phase was (A) water with formic acid 0.1% and

(B) acetonitrile, the flow rate was 0.3 mL/min and the elution program from 20% (B) to 40 % (B) in 5 min. The LC system was coupled to a triple-stage quadrupole mass spectrometer (Applied Biosystem/5500 QTrap) with electrospray ion source in positive mode for the analysis. The turboionspray source parameters were optimized in positive mode and were as follows: ion spray voltage 5500 V, curtain gas 20 (arbitrary units), nebulizer gas 1 and auxiliary gas 2 respectively 50 and 60, probe temperature of 450°C.

### 2.6. Statistical analysis

Data are reported as mean ± standard deviation (SD) and statistical analysis was performed using SPSS 17.0 software. Firstly, the normal distribution and homoscedasticity of data were verified using Kolmogorov-Smirnov and Levene tests respectively ($\alpha=0.05$). A two-way analysis of variance (ANOVA) was performed to assess gender effect on KEH values in bullhead. In the present work, no significant difference was observed, and male and female KEH values were pooled. Differences in KEH values between exposure conditions and durations were analyzed by a Mann-Witney $U$ test ($\alpha=0.05$).

### 3. Results

### 3.1 Location and structure of bullhead kidney

In the bullhead, the kidney is located retroperitoneally, outside the peritoneal cavity (Figure 1a). Macroscopic observation of the kidney shows that the head kidney is constituted by bilateral lobes separated from the trunk kidney by two slender extensions located on both sides of vertebral column (Figure 1a, b). Histological analysis reveals a lack of nephritic tissue in this section and only haematopoietic tissues are observed (Fig. 1c). The posterior kidney of the bullhead was elongated and was composed of nephritic tissue including proximal and distal tubules, renal corpuscles and collecting ducts (25 ± 5%) and haematopoietic tissues (75 ± 5 %) (Fig. 1d).

### 3.2. Natural variation of bullhead kidney epithelium height

Histological observation of the renal tissue of male and female bullheads randomly electrofished in an uncontaminated river showed clear sex-dependent hypertrophy during the breeding season (Figure 2). The results of KEH measurement are presented in Figure 3 and confirm this hypertrophy. The mean value of KEH in male bullheads is 20.96 ± 3.57 µm whereas it is 16.96 ± 1.28 µm for females during the breeding season. Outside of the breeding period, sexual dimorphism was not observed based on KEH measurement. Mean KEH values are 16.49 ± 1.19 µm and 15.79 ± 0.86 µm for male and female bullheads respectively.

### 3.3. Waterborne exposure to trenbolone acetate

Measurement of TbA in the experimental water confirmed that control fish were not exposed to this chemical. The time-weighted mean measured concentrations of TbA throughout the 21-day exposure period showed that fish were exposed to 1.26 and 6.50 µg/L (Table 1). After trenbolone exposure, histological observation of the kidney's ultra-structure showed clear dose-dependent hypertrophy based on KEH measurement (Figure 4). DMSO-exposed bullheads presented basal KEH values of 14.58 ± 0.85 µm and 12.65 ± 1.03 µm after 14 and 21 days respectively. A significant dose-dependent induction was noticed in bullheads exposed to 1.26 µg/L of TbA (KEH values of 16.89 ± 0.81 µm and 17.73 ± 0.52 µm for 14 and 21 days respectively) and 6.50 µg/L of TbA (KEH values of 21.01 ± 1.76 µm and 19.33 ± 1.03 µm for 14 and 21 days respectively). The maximal induction was observed for bullheads exposed to TbA at 6.50 µg/L for 14 days with an induction factor of 1.4.

### 4. Discussion

The major aim of the present study was to identify a potential biomarker of androgen exposure in the kidney of the European bullhead. Indeed, the three-spined stickleback is currently the only fish species able to provide data on fish exposure to androgenic compounds in European water bodies, due to the existence of biochemical and histological specific biomarkers linked to spiggin synthesis (Katsiadaki et al. 2002; Sanchez et al. 2008a). However, this fish species, described as a relevant sentinel species for freshwater biomonitoring (Sanchez et al. 2008b), is not widely distributed in European streams particularly in large rivers due to the lack of specific habitats (Sanchez et al. 2007). Histological studies performed in the kidney of the European bullhead revealed a gender-dependent hypertrophy with an increase in nephritic channel size in breeding males (Bucher and Hofer 1993; Hentschel 1982). This phenomenon is described as comparable to the natural change in stickleback kidney ultra-structure associated with spiggin synthesis during the breeding season and could be a potential biomarker of androgen exposure. Combined with the wide distribution of the European bullhead in European countries (Freyhof et al. 2005) and the potential of this fish species for environmental toxicology studies in field and laboratory conditions (Jolly et al. 2012; Dorts et al. 2011a; Dorts et al. 2011b), the hypothesis of a potential biomarker of androgen exposure offers interesting perspectives for the use of the bullhead as a relevant sentinel fish species in monitoring studies.

Firstly, the structure of the bullhead's kidney was documented to identify the normal structure of this organ. As observed in other teleost fish, the histology of the European bullhead's kidney showed two separate extensions in the most anterior part of the organ (Press and Evensen 1999; Grassi Milano et al. 1997) which form a separation between anterior aglomerular (head kidney) and mid and posterior glomerular (trunk kidney) compartments (Press and Evensen 1999). Currently, these two compartments possess various anatomies and functionalities which could vary between species. Concerning the head kidney, as for other fish species, this compartment is lacking in nephrons and largely composed of haematopoietic and lymphopoietic tissues. The trunk kidney of the bullhead had few nephrons (25 ± 5%) and much immune tissue (75 ± 5 %) which was not limited to the area around the renal tubules

(Lin et al. 2005; Whyte 2007). This observation is not in accordance with histological descriptions of the trunk kidney structure in other fish species such as the stickleback, which exhibits haematopoetic tissue and is located in the interstitial region surrounding renal tubules (Mourier 1970). So, in the European bullhead, the excretion function is not observed in the cephalic part of the kidney but both renal regions exhibit haematopoietic and lymphopoietic capacities.

Secondly, sexual and seasonal changes to the structure of the trunk kidney were assessed and quantified. For this purpose, the measurement of KEH was used. This tool was previously developed and validated for stickleback to quantify the enlargement of secondary proximal tubules due to spiggin synthesis during the reproductive period in males or after exposure to androgen (Borg 1993). This quantitative approach was applied to the kidney of the European bullhead and confirms the presence of hypertrophy of secondary proximal tubules in male bullheads during the reproductive period as previously described by (Bucher and Hofer 1993). These authors compared the hypertrophy observed in the kidney of the bullhead to that reported in the stickleback, explained by spiggin synthesis. In the bullhead, the molecular mechanisms supporting this ultrastructural modification are unknown. (Hentschel 1982) linked kidney hypertrophy to the production of sialoglycoprotein-containing mucus in the epithelium of the second proximal segment of the kidney during the breeding period but these proteins and associated functions are not identified. However, due to strong similarity between bullhead and stickleback kidneys, the effect of androgen exposure on the ultrastructure of the bullhead kidney was investigated. Indeed, spiggin is an androgeno-regulated protein and physiological changes associated with spiggin induction can be explained by a high concentration of 11-ketotestosterone prior to the breeding period (Jakobsson et al. 1999).

In a last step, adult bullheads were waterborne exposed to TbA, a synthetic growth promoter in beef cattle recognized as a powerful androgen able to masculinize a zebrafish population at environmentally relevant concentrations (Morthorst et al. 2010). Trenbolone, the major metabolite of TbA, is also recognised as a powerful androgen known to induce spiggin after 14 days of exposure to 5 µg/L (Allen et al. 2008). In the present work, TbA induced ultrastructural changes in the kidney of bullheads with an enlargement of proximal tubules. This effect is quite similar to the seasonal change observed in male bullheads during the breeding period. A 1.3-fold induction was noticed in breeding males and a 1.4-fold induction was reported after 14 days of exposure to 6.5 µg/L of TbA. However, compared to KEH in stickleback, this end-point appears as less sensitive. For example, 2.0- and 2.3-fold inductions were measured in castrated sticklebacks exposed respectively to 25 µg trenbolone/g or a 5 µg 11β-hydroxyandrostenedione/g (Borg 1993). Similar data was obtained in the same conditions as bullheads. Moreover, a 2.1-fold induction was observed in sticklebacks waterborne exposed to 6.5 µg TbA/L for 14 days of exposure (data not shown). These results obtained with the stickleback are consistent with data available in the scientific literature (Borg 1993) and confirm the lower sensitivity of KEH response in the bullhead. Also, due to a weak induction factor, KEH measurement in the bullhead will be less discriminating than in the stickleback.

To summarize, the data presented in the present paper shows that the kidney of the bullhead exhibits a gender and seasonal variation of the ultrastructure that can be induced by TbA, a model androgen. These changes can be measured using the KEH previously developed in the stickleback. However, KEH is a time-consuming method (Katsiadaki et al. 2002) and appears, in the bullhead, as a less sensitive end-point to quantify androgen exposure. Hence, further studies are needed to identify molecular signals that could be more sensitive than KEH and could be used in laboratory ecotoxicological assays and in environmental monitoring programs.

**Acknowledgements**

This work was funded by the European Regional Development Fund in the framework of the INTERREG IV A France (Channel) - England program (DIESE project) and by the French Ministry for Ecology (Programme 190-Ecotoxicology).

**Figure captions**

Figure 1. Photograph showing the kidney location in European bullhead (*Cottus* sp.) (a) and external view of anterior (arrow) and trunk kidney (TK) (b). Scale bar = 0.5 cm. Transverse section of European bullhead head kidney, Bouin, HE, x 100 (c). Transverse section of European bullhead trunk kidney, Bouin, HE, x 100 (d). 1. Glomerulus, 2. Renal tubule, 3. Lymphoid tissue.

Figure 2. Kidney sections of female and male bullhead (*Cottus* sp.) in breeding season (late March) or in non breeding season (early July). Sections were stained with Haematoxylin-Eosin. Presented microphotographs were taken under the same magnification and the scale bar represents 30 μm.

Figure 3. Kidney Epithelium Height (KEH) values of male and female bullheads *(Cottus sp.)* in breeding season or in non breeding season. Data are presented as mean ± SD and the number refers to the number of fish sampled. For both sexes, histograms with the same letter are not statistically different ($p<0.05$).

Figure 4. Kidney Epithelium Height (KEH) values of bullheads *(Cottus sp.)* waterborne exposed for 14 (black bars) and 21 days (grey bars) to two concentrations of trenbolone acetate. Data are presented as mean ± SD and the number refers to the number of fish in each treatment. For both exposure times, histograms with the same letter are not statistically different ($p<0.05$).

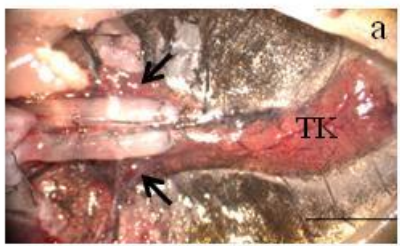

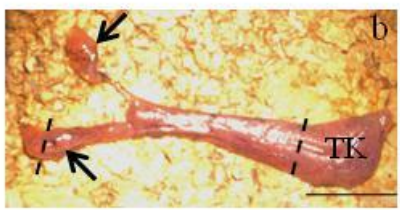

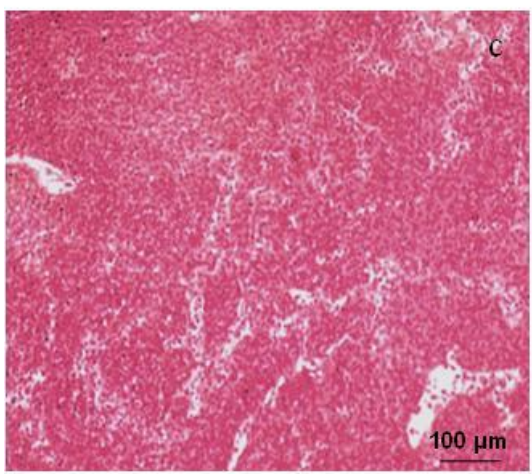
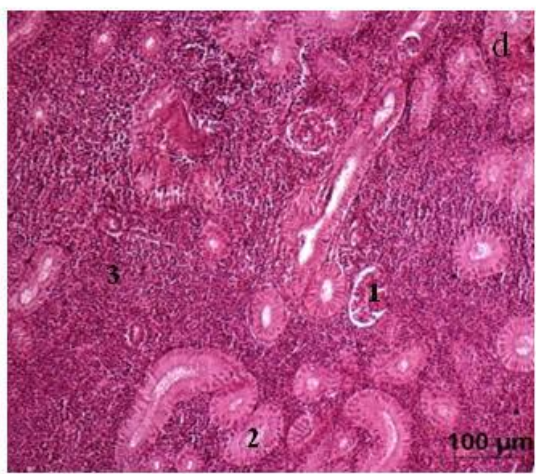

Figure 1

|  | female | male |
|---|---|---|
| **Breeding** | 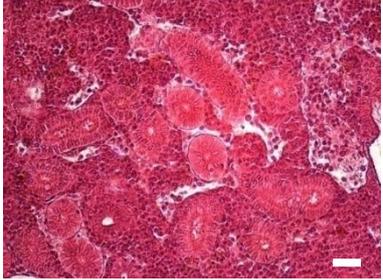 | 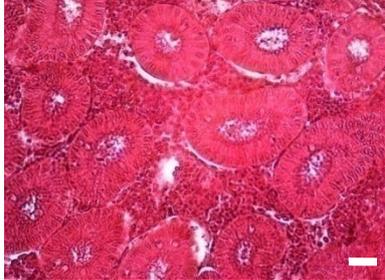 |
| **Non breeding** | 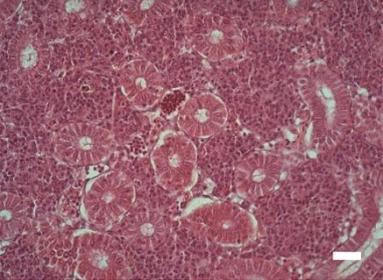 | 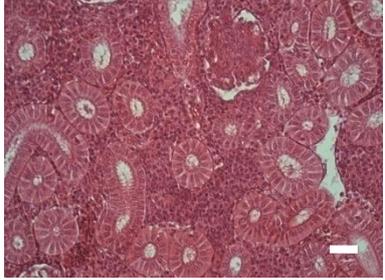 |

Figure 2

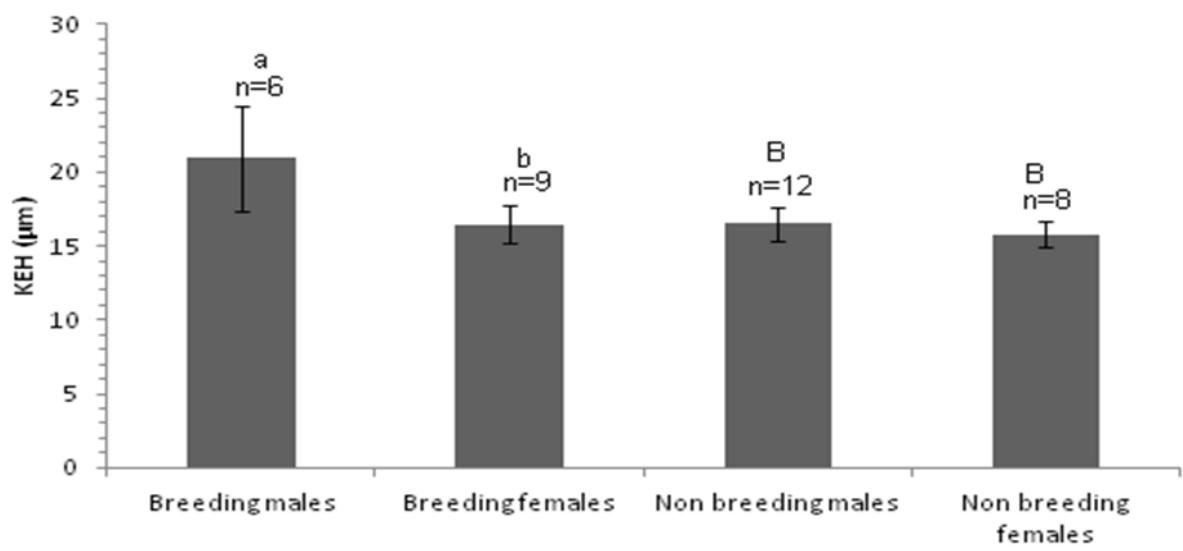

Figure 3

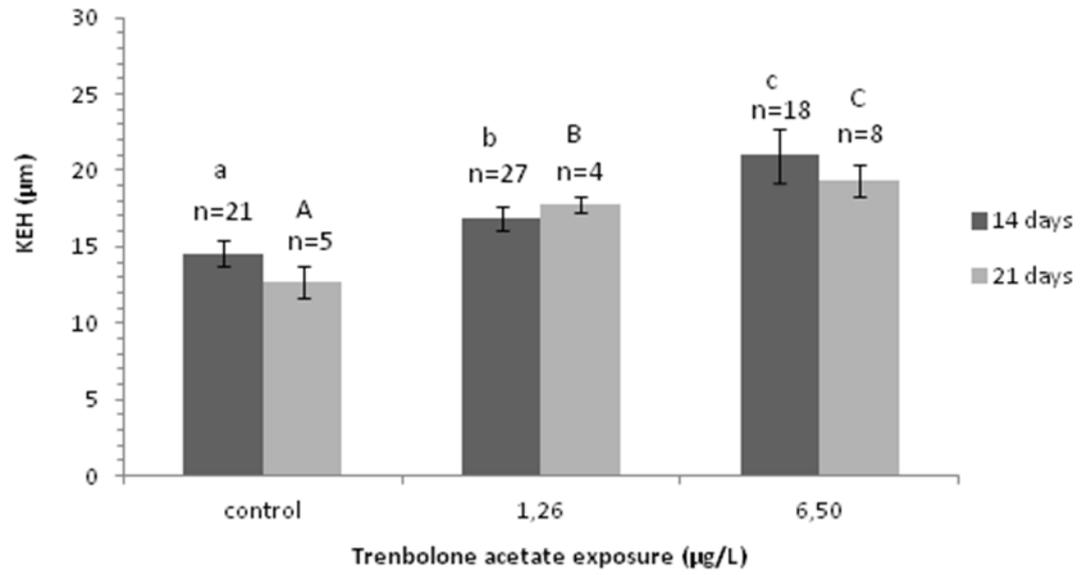

Figure 4

Table 1. Analytical measurement of Trenbolone acetate concentrations in water of exposure tank.

|  | Nominal concentration | Measured concentration | Limit of Detection (LD) | Limit of quantification (LOQ) |
|---|---|---|---|---|
| Control | 0 µg/L | ND | 0.001 µg/L | 0.05 µg/L |
| TbA | 0,5 µg/L | 1.26 µg/L | | |
|  | 5 µg/L | 6.50 µg/L | | |

ND: no detectable concentration